\documentclass{iaus}
\usepackage[dvips]{graphicx}
\usepackage[T1]{fontenc}
  \checkfont{eurm10}
  \iffontfound
    \IfFileExists{upmath.sty}
      {\typeout{^^JFound AMS Euler Roman fonts on the system,
                   using the 'upmath' package.^^J}%
       \usepackage{upmath}}
      {\typeout{^^JFound AMS Euler Roman fonts on the system, but you
                   dont seem to have the}%
       \typeout{'upmath' package installed. iaus.cls can take advantage
                 of these fonts,^^Jif you use 'upmath' package.^^J}%
      }
  \else
  \fi

  \checkfont{msam10}
  \iffontfound
    \IfFileExists{amssymb.sty}
      {\typeout{^^JFound AMS Symbol fonts on the system, using the
                'amssymb' package.^^J}%
       \usepackage{amssymb}%

      }{}
  \fi

  \IfFileExists{amsbsy.sty}
    {\typeout{^^JFound the 'amsbsy' package on the system, using it.^^J}%
     \usepackage{amsbsy}}
    {}

\newsavebox{\astrutbox}
\sbox{\astrutbox}{\rule[-5pt]{0pt}{20pt}}

\newcommand\etal{\mbox{\textit{et al.}}}

\newcommand{\ngc}{NGC\,1097}

\newcommand{\qm}[1]
{``#1''}

\newcommand{\eb}{\qm{energy budget} }

\title[The Interplay among Black Holes, Stars and ISM in Galactic
       Nuclei]{The Ionizing Source of the Nucleus of NGC1097}

\author[R. S. Nemmen {\it et al.\/}]%
{Rodrigo S. Nemmen$^1$,
Thaisa Storchi-Bergmann$^1$, \break Michael Eracleous$^2$,
Yuichi Terashima$^3$ \and Andrew Wilson$^4$}

\affiliation{$^1$Instituto de Física, UFRGS, Porto Alegre, Brasil \\[\affilskip]
$^2$ Department of Astronomy and Astrophysics, Penn State University, USA \\[\affilskip] $^3$ Institute of Space and Astronautical Science, Japan \\[\affilskip] $^4$ Astronomy Department, University of Maryland, USA}

\pubyear{2004}
\volume{222}
\pagerange{1--8}
\date{?? and in revised form ??}
\jname{The Interplay among Black Holes, Stars and ISM \\in Galactic Nuclei}
\editors{Th. Storchi Bergmann, L.C. Ho \& H.R. Schmitt, eds.}
\begin{document}

\maketitle

\begin{abstract}
We present new observations in X-ray and optical/ultraviolet of the nucleus of \ngc, known for the abrupt appearance of broad, double-peaked Balmer lines in its spectrum in 1991. These new observations are used to construct the spectral energy distribution (SED) of the central engine. From the SED we infer that this AGN is radio-loud and has a bolometric luminosity $L_{\rm Bol} \sim 10^{42} \: \textrm{erg s}^{-1}$, implying a low Eddington ratio of $L_{\rm Bol}/L_{\rm Edd} \sim 10^{-4}$. These results suggest that the central ionizing source is an advection-dominated accretion flow (ADAF) in the form of an ellevated structure which photoionizes an outer thin disk. We fit a simplified ADAF model to the SED and obtain limits on the values of the mass accretion rate $\dot{M}$ and accretion efficiency $\eta$, namely $\dot{M}/{\dot{M}_{\rm Edd}} \gtrsim 10^{-3}$ and $\eta \lesssim 10^{-2}$. We identify an \eb problem: if the central photoionizing source is isotropic, the covering factor of the line-emitting portion of the thin accretion disk is $\approx 6$, i. e. the central source accounts for only 20\% of the energy emitted in the double-peaked Balmer lines.
\end{abstract}

\firstsection % if your document starts with a section,
              % remove some space above using this command.
\section{Introduction}

\ngc \ is a spiral galaxy which harbors a Seyfert 1/LINER active nucleus (AGN). This AGN emits a broad double peaked H$\alpha$ emission-line (\cite[Storchi-Bergmann \etal\ 1993]{sb93}\footnote{See also contributions of Eracleous \etal, Lewis \etal\ and Gezari \etal\ in this volume, concerning douple-peaked emitters}), which is interpreted as the kinematic signature of a relativistic accretion disk around the central black hole (\cite[Storchi-Bergmann \etal\ 2003]{sb03}, hereafter SB03). This accretion disk was probably formed from the tidal debris of a star disrupted by a supermassive black hole\footnote{See also contributions of Komossa et al. and Bogdanovic et al. in this volume, concerning tidal disruption of stars}. In this work we take a step further in the study of the central engine of this galaxy, trying to understand the ionizing source which is supposed to be the origin of the radiation that illuminates the  accretion disk. Towards this goal, we present new observations of the nuclear continuum in X-rays (Chandra) and optical/ultraviolet (HST), which are used to construct the spectral energy distribution (SED) of the nucleus. We model this SED using a simplified advection-dominated accretion flow (ADAF) prescription (see review by \cite[Narayan \etal\ 1998]{narayan98}) which allows us to obtain some basic quantitative information about the accretion process, such as limits on the mass accretion rate and the efficiency of conversion of gravitational energy into radiation.

The high resolution SED presented in this work, combined with observations of the broad H$\alpha$ profile from SB03 allows us also to address the energetics of the thin accretion disk and the ion torus.

\section{The Spectral Energy Distribution}\label{sec:sed}

\begin{figure}
\centering
\includegraphics[scale=0.6]{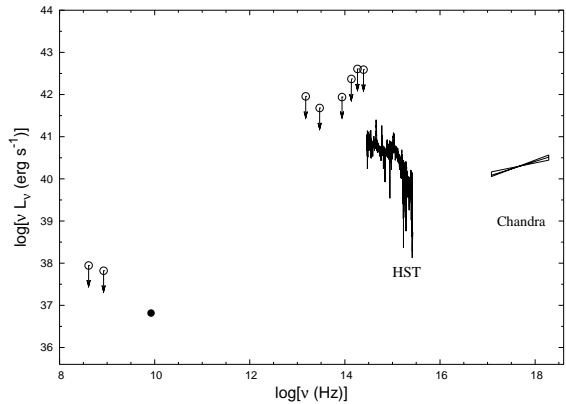}
\caption{SED of the nucleus of \ngc. The open circles denote observations which include contamination by the host galaxy and are thus upper limits. The filled circle is a radio point dominated by the nuclear continuum.}
\label{fig:sed}
\end{figure}

We observed the nucleus of \ngc\ with Chandra using an aperture of 2.6 arcsec, and with STIS on HST using various gratings and an aperture of $0.2 \times 0.2 \: \textrm{arcsec}^2$. We also collected radio and infrared data from the literature, and the resulting SED is shown in Figure \ref{fig:sed}.

%\begin{figure}
%\begin{minipage}[t]{0.48\linewidth}
%\includegraphics[width=\linewidth]{SED.eps}
%\caption{SED of the nucleus of \ngc. The open circles denote observations which include too much contamination by the host galaxy. The filled circle is a radio point dominated by the nuclear continuum.}
%\label{fig:sed}
%\end{minipage} \hfill
%\begin{minipage}[t]{0.48\linewidth}
%\includegraphics[width=\linewidth]{comp.eps}
%\caption{Comparison of the nuclear SED with those of other low-luminosity AGNs (dotted line), median radio-loud (dashed line) and radio-quiet (dot-dashed line) quasars. The spectra are normalized such that their luminosity at 1 keV is similar \naosei{equal}.}
%\label{fig:comp}
%\end{minipage} \hfill
%\end{figure}

\begin{figure}
\centering
\includegraphics[scale=0.6]{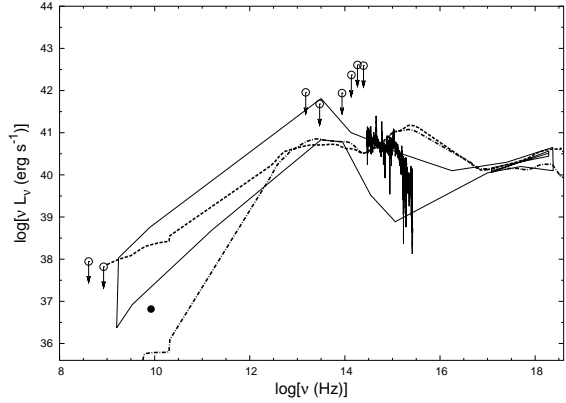}
\caption{Comparison of the nuclear SED with those of other low-luminosity AGNs (solid line), median radio-loud (dashed line) and radio-quiet (dot-dashed line) quasars. The spectra are normalized such that their luminosities at 1 keV are similar.}
\label{fig:comp}
\end{figure}

It is interesting to compare the continuum of \ngc\ with those of other low-luminosity AGNs (hereafter LLAGNs). The dotted lines of Figure \ref{fig:comp} enclose the region covered by the LLAGN SEDs studied by \cite{ho99} from which we conclude that \ngc\ 's SED is similar to those of other LLAGNs. The observed bolometric luminosity is $L_{\rm Bol} = 1.2 \times 10^{42} \; \textrm{erg s}^{-1}$, implying a low Eddington ratio of $\frac{L_{\rm Bol}}{L_{\rm Edd}} \sim 10^{-4}$ \footnote{The adopted black hole mass is $5 \times 10^{7} M_{\odot}$ (SB03).} and a ratio of X-ray luminosity (0.5 - 10 keV) to $L_{\rm Bol}$ of 0.054. This AGN is marginally radio-loud according to both the
optical criterion
($R_{o} = \frac{F_{\nu} (6 \; \mathrm{cm})}{F_{\nu} (B)} = 14$; the boundary
between radio-loud and radio-quiet is
$R_{o} = 10$) and the X-ray criterion
($R_{\rm X} = \frac{\nu L_{\nu}(5 \; \mathrm{GHz})}{L_{\rm X}} = 1.4 \times 10^{-4}$; here
L$_{\rm X}$ is the luminosity in the 2 - 10 keV band; the boundary is
3.2 $\times$ 10$^{-5}$)
defining radio loudness (\cite[Terashima \& Wilson 2003]{terashima03}), much like other LLAGNs (\cite[Ho 1999]{ho99}, \cite[Ho \etal\ 2000]{ho00}).
%\naosei{The definition of radio loudness using
%the hard X-ray flux (R$_{\rm X}$) is more robust than the optical definition
%because hard
%X-rays are much less affected by extinction than are optical wavelengths.
%Furthermore, measurements of nuclear X-ray fluxes of Seyferts and LINERS with
%Chandra are easier than measurements of nuclear optical fluxes, since in
%the latter case the nuclear light must be separated from the surrounding
%starlight, a difficult process when the nucleus has low luminosity.}

Figure \ref{fig:comp} also shows a comparison of \ngc\ 's SED with median spectra of well-studied radio-loud and radio-quiet quasars (\cite[Elvis \etal\ 1994]{elvis94}), from which we see that the continuum of \ngc\ lacks the classical UV bump in the range of energies common to quasars (\cite[Ho \etal\ 2000]{ho00}), indicating that the thin accretion disk in this case is probably cooler than the disks of quasars (bump is located at lower energies).

\section{Modelling} \label{sec:model}

\begin{figure}
\centering
\includegraphics[scale=0.6]{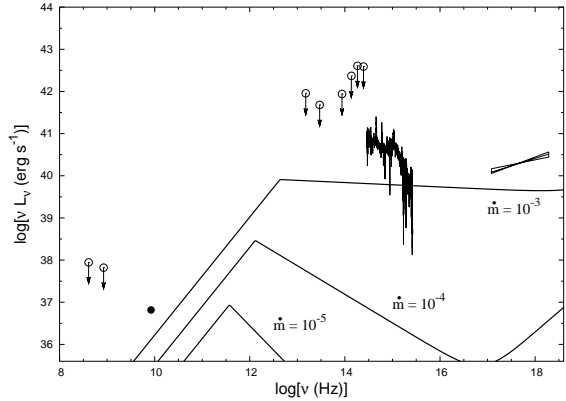}
\caption{Theoretical spectra from ADAF compared with the observations. The solid lines represent ADAF models generated with different values for the accretion rate.}
\label{fig:adaf}
\end{figure}

An appealing scenario for explaining the observational results mentioned above is that the continuum emission originates in an ADAF, in the form of a geometrically thick ion-supported torus that photoionizes the outer thin accretion disk. We adopted the simplest available ADAF model (\cite[Mahadevan 1997]{mahadevan97}), which can be thought of as a zeroth order approximation to the radiative processes. This model has the advantage of providing analytical expressions for the SED. The model parameters are the mass accretion rate, $\dot{M}$, the black hole mass, $M$, the inner and outer boundaries of the flow, $R_{\rm in}$ and $R_{\rm out}$, the ratio of gas pressure to total pressure, $\beta$, and the viscosity parameter, $\alpha$.
The estimated parameters are: $M=(2-9) \times 10^7 \: M_{\odot}$ (SB03), $R_{\rm in}=3 \: R_{\rm Schw}$, $R_{\rm out}=225 \: R_{\rm Schw}$ (SB03), where $R_{\rm Schw}$ is the Schwarzschild radius; $\beta$ and $\alpha$ are kept at their typical values, respectively 0.5 and 0.1. The only free parameter that remains is $\dot{M}$. We find a lower limit of $\dot{M} \gtrsim 10^{-3} \: M_{\odot} \: \textrm{year}^{-1}$, which translates to $\dot{m}=\frac{\dot{M}}{\dot{M}_{\rm Edd}} \gtrsim 10^{-3}$, where $\dot{M}_{\rm Edd}=\frac{L_{\rm Edd}}{0.1 c^2}$ is the Eddington accretion rate. Because $\dot{M}=\frac{L_{\rm Bol}}{\eta c^2}$, with $\eta$ the efficiency of conversion of gravitational energy into radiation, the lower limit of $\dot{M}$ imposes an upper limit on the efficiency such that $\eta \lesssim 10^{-2}$. Figure \ref{fig:adaf} shows some theoretical spectra for the ADAF. Due to its simplicity, the model cannot account for the fine details of the SED; it can only reproduce the overall luminosity of the continuum if $\dot{m} \gtrsim 10^{-3}$.

\section{Photon Balance}

We also calculated the photon balance between the central source and the thin accretion disk. From the broad H$\alpha$ emission-line we computed the total number of recombination photons originating in the outer accretion disk, which is $N_{\rm rec} \sim 10^{-2}$ photons cm$^{-2}$ s$^{-1}$. The number of ionizing photons from the nuclear continuum is $N_{\rm ion} \sim 10^{-3}$ photons cm$^{-2}$ s$^{-1}$. Assuming that photoionization by an isotropic source powers the emission-lines, we would expect that $\frac{N_{\rm rec}}{N_{\rm ion}} \lesssim 1$ (a covering factor smaller than unity), but the ratio that we get is 6. This ratio shows that the photoionizing radiation from the source most probably accounts for only 20\% of the recombination photons emitted by the outer thin disk. This seems to be a manifestation of the more general \eb problem, which states that the central engine is not energetic enough to power the emission-lines of the broad and narrow line regions. This inconsistency in the photoionization scenario was first discussed in the context of quasars by \cite{netzer85}, and it still seems to be an unsolved problem in the study of AGNs.

There may be some explanations to this discrepancy:
\begin{itemize}
    \item Additional sources of ionization of the gas beyond photoionization, for example shock heating or dissipative turbulent heating (\cite[Bottorff \& Ferland 2002]{bottorff02});
    \item The continuum is emitting anisotropically such that we don't see the same continuum \qm{seen} by the outer accretion disk (\cite[Korista \etal\ 1997]{korista97});
    \item Additional continuum reddening, besides the Milky Way foreground reddening used to correct the UV-optical data and that derived from the hydrogen column density obtained from X-ray data, which is very similar to the Milky Way one, which has already been taken into account.
\end{itemize}

\section{Conclusions}

We have presented the nuclear SED of \ngc, concluding that the nucleus is radio loud ($R=14$). Its bolometric luminosity of $L_{\rm Bol} = 1.2 \times 10^{42} \; \textrm{erg s}^{-1}$ implies $\frac{L_{\rm Bol}}{L_{\rm Edd}} \sim 10^{-4}$ and $\frac{L_{\rm X}}{L_{\rm Bol}} = 0.054$. The SED is similar to those of other LLAGNs, but lacks the classical UV bump in the range of energies common to quasars, probably indicating that the thin accretion disk is cooler than typical disks of quasars. Modelling the SED using a simplified ADAF prescription, we find that $\dot{M} \gtrsim 10^{-3} \: M_{\odot} \: \textrm{year}^{-1}$, such that $\dot{m} \gtrsim 10^{-3}$ and $\eta \lesssim 10^{-2}$. We obtain a covering factor for the line-emitting disk of $\frac{N_{\rm rec}}{N_{\rm ion}} \approx 6$, suggesting an \eb problem.

A more complete version of this work, including detailed modelling of the SED and further investigation of the \eb problem will be presented elsewhere.

\begin{acknowledgments}
This work was partially supported by the Brazilian institutions CAPES and CNPq, and by NASA through grant NAG 8-1755 to the University of Maryland.
\end{acknowledgments}


\begin{thebibliography}{}

\bibitem[Bottorff \& Ferland (2002)]{bottorff02} Bottorff, M., \& Ferland, G. 2002, ApJ, 568, 581
%\bibitem[Dumont, Collin-Souffrin \& Nazarova (1998)]{dumont98} Dumont, A-M., Collin-Souffrin, S., \& Nazarova, L. 1998, A \& A, 331, 11
\bibitem[Elvis (1994)]{elvis94} Elvis, M., \etal\ 1994, ApJS, 95, 1
\bibitem[Ho (1999)]{ho99} Ho, L. 1999, ApJ, 516, 672
\bibitem[Ho \etal\ (2000)]{ho00} Ho, L., \etal\ 2000, ApJ, 541, 120
\bibitem[Korista, Ferland \& Baldwin (1997)]{korista97} Korista, K., Ferland, G., \& Baldwin, J. 1997, ApJ, 487, 555
\bibitem[Mahadevan (1997)]{mahadevan97} Mahadevan, R. 1997, ApJ, 477, 585
\bibitem[Narayan, Mahadevan \& Quataert (1998)]{narayan98} Narayan, R., Mahadevan, R., \& Quataert, E. 1998, in The Theory of Black Hole Accretion Disks, ed. M. A. Abramowicz, G. Bj\"ornsson, \& J. E. Pringle (Cambridge: CUP), 148
\bibitem[Netzer (1985)]{netzer85} Netzer, H. 1985, ApJ, 289, 451
\bibitem[Storchi-Bergmann, Baldwin \& Wilson(1993)]{sb93} Storchi-Bergmann, T., Baldwin, J. A., \& Wilson, A. S. 1993, ApJ, 410, L11
\bibitem[Storchi-Bergmann \etal\ (2003)]{sb03} Storchi-Bergmann, T., et al. 2003, ApJ, 598, 956 (SB03)
\bibitem[Terashima \& Wilson (2003)]{terashima03} Terashima, Y., \& Wilson, A. 2003, ApJ, 583, 145

\end{thebibliography}
\end{document}